\documentclass[12pt]{article}
\usepackage{amsmath}
\usepackage{amstext}
\usepackage{amsfonts}
\usepackage{amsbsy}
\usepackage{eucal}
\usepackage{amssymb}
\usepackage{amsthm}
\usepackage{color}
\usepackage[dvips]{graphicx}
\newtheorem{theorem}{Theorem}[section]

\theoremstyle{definition}

\newtheorem{remark}[theorem]{Remark}
\newcommand{\ov}{\overline}
\newcommand{\un}{\underline}
\newcommand{\wt}{\widetilde}
\begin{document}
\begin{center}
{\bf Multisoliton solution and super-bilinear form of lattice supersymmetric KdV equation}
\end{center}
\begin{center}
A. S. Carstea
\end{center}
\begin{center}
{\it Dept. of Theoretical Physics, Institute of Physics and Nuclear Engineering, 407 Atomistilor, Magurele, 077125 Bucharest, Romania}\\
\end{center}

\begin{abstract}
Hirota bilinear form and multisoliton solution for semidiscrete and fully discrete (difference-difference) versions of supersymmetric KdV equation found by Xue, Levi and Liu \cite{liu} is presented. The solitonic interaction term displays a fermionic dressing factor as in the continuous supersymmetric case. Using bilinear equations it is shown also that there can be constructed  a new integrable semidiscrete (and fully discrete) version of supersymmetric KdV which has simpler bilinear form but more complicated interaction dressing. Its continuum limit is also computed.

\end{abstract}

\section{Introduction}

Supersymmetric integrability is a fascinating topic since super-extensions of nonlinear systems have richer phenomenolgy and still not all the tools used in classical integrability are effective in supersymmetric context. Indeed Lax/hamiltonian integrability works perfectly in the SUSY context (as well as prolongation structure, Painleve test etc; see \cite{aratyn} and references therein) but bilinear formulation of super-integrable hierarchies was not very deep investigated, although the nonlinear formulation was done a long time ago by Manin  and Radul \cite{manin}
The basic idea in the bilinear formulation is that solitons in a hierarchy are behaving like free fermions in a conformal field theory on Riemann surface. In the case of supersymmetric solitons this fact is not quite clear although many results have been found. We mention here the bilinear formulation of Kac and van de Leur \cite{kac1}, Ueno-Yamada-Ikeda \cite{ueno}, Kac and Medina \cite{kac2}, Le Clair \cite{leclair} but it is quite hard to get the nonlinear forms and solutions out of them. 
The direct bilinear approach on equations (not hierachies) was done in \cite{carstea0} where, after construction of the super-extension of Hirota operator, we studied various equations (KdV, SG, SK, Painleve I, II and some linearisable systems - superRicatti, superLiouville and superBurgers ) \cite{carstea1, carstea2, carstea3, carstea4}. A quite nice property was found at the construction of multisoliton solution, namely the presence of a fermionic dressing at the interaction of supersolitons. This fact makes the analysis of super-soliton interaction quite hard. In addition, many other interesting properties appeared, for instance the Burgers equation admits two super-extensions one being linearisable and the other bilinearisable \cite{carstea3}. Also fermionic/supersymmetric extensions of Painleve I and II equations have many unusual properties which deserve further investigations \cite{carstea4}.
After that many other integrable supersymmetric extensions have been bilinearized like mKdV, two-boson system etc \cite{hu1,hu2,c1,c2}. Bilinear forms and soliton solutions for fermionic non-supersymmetric extension were  very little investigated. We obtained some results related to fermioninc extensions of Zakharov and Yajima-Oikawa systems in \cite{ssh1,ssh2} 

The case of discrete equations seems to be even harder. The nonlocality of the lattice makes the supersymmetry implementation to be very difficult so a direct approach must be done. In the case of discrete Toda lattice and discrete nonlinear Schr\"odinger, Grahovski and Mikhailov \cite{grakhovski} succeded in finding an integrable  super-extension. The case of discrete KdV equation has been analysed recently by Xue, Levi and Liu and they found interesting integrable super extensions of potential discrete KdV \cite{liu} namely a semidiscrete one:
\begin{equation}\label{liu1}
\partial_t \psi_n=2\frac{\psi_{n+1}-\psi_{n-1}}{u_{n+1}-u_{n-1}-4p}
\end{equation}
\begin{equation}\label{liu2}
\partial_t u_n=2\frac{u_{n+1}-u_{n-1}}{u_{n+1}-u_{n-1}-4p}+\frac{u_{n+1}-u_{n-1}-8p}{(u_{n+1}-u_{n-1}-4p)^2}(\psi_{n+1}-\psi_{n-1})(\psi_n-\psi_{n-1})
\end{equation}
and fully discrete one:
\begin{equation}\label{diliu1}
\psi_{n+1,m+1}-\psi_{n,m}=\frac{2(p_1+p_2)(\psi_{n+1,m}-\psi_{n,m+1})}{2(p_2-p_1)+u_{n+1,m}-u_{n,m+1}}
\end{equation}
$$
u_{n+1,m+1}-u_{n,m}=\frac{2(p_1+p_2)(u_{n+1,m}-u_{n,m+1})}{2(p_2-p_1)+u_{n+1,m}-u_{n,m+1}}-$$
\begin{equation}\label{diliu2}
-\frac{(p_1+p_2)(4(p_2-p_1)+u_{n+1,m}-u_{n,m+1})}{(2(p_2-p_1)+u_{n+1,m}-u_{n,m+1})^2}(\psi_{n+1,m}-\psi_{n,m+1})(\psi_{n,m+1}-\psi_{n,m})
\end{equation}
Here $\psi_n$ and $u_n$ are the fermionic and bosonic fields dependind on $n$ and $t$ (or $n$ and $m$ respectively)with values in the odd and even sector of an infinite dimensional Grassmann algebra. Integrability of the above system was established by displaying the B\"acklund transformations and Lax pair.

However because the construction of multisoliton solution was not done, we intend in this paper to formulate the Hirota form and compute the super-soliton solution. Also, starting from the differential-difference super-bilinear equations, we will construct  fully discretized bilinear system which turns into the nonlinear form found by Xue, Levi and Liu. In addition, we shall provide a new form of discrete super-KdV equation with a simpler bilinear structure but with a more complicated soliton interaction. Its continuum limit goes to a fermionic extension of KdV which after a simple transformation is equivalent to supersymmetric KdV equation.
The paper is organised as follows; in the section 2 we give a brief description of the bilinear approach potential (semi)discrete KdV equation. We will show how to construct soliton solution and how to discretize the bilinear form in order to obtain the fully discretized version (giving an alternative construction of the bilinear form of Hietarinta and Zhang \cite{jarmo} for H1 equation). Then we will remind the bilinear formulation of supersymmetric KdV equation focusing on the component rather than superspace description. In section 3 and 4 we bilinearize and compute multisoliton solutions for the semidiscrete and fully discrete versions and analyse the continuum limit. 
\section{Discrete KdV and continuous super KdV}
The semidiscrete version of KdV equation has been obtained long time ago and has the following expression \cite{hirota}:
\begin{equation}\label{sk1}
u_{n,t}=u_n^2(u_{n+1}-u_{n-1}).
\end{equation}
The potential Kac-van Moerbecke form of the semidiscrete KdV can be obtained by taking $u_n(t)=\partial_t w_{n}(t)-1/4p$ ($p$ is an arbitrary constant), integrating once, and take the constant of integration to be also $4p$. The equation (\ref{sk1}) will turn into:
\begin{equation}\label{kac1}
w_{n,t}=\frac{w_{n+1}-w_{n-1}}{w_{n+1}-w_{n-1}-4p}.
\end{equation}
The constant $4p$ can be rescaled to 1 by taking $w_n\to 4p w_n, \partial_t\to4p\partial_t$.
Solutions can be found using the bilinear formalism. Indeed if $w_n=g_n/f_n$ we will get:
\begin{equation}\label{bil0}
D_tg_n\cdot f_n=(g_{n+1}f_{n-1}-g_{n-1}f_{n+1}),
\end{equation}
\begin{equation}\label{bil1}
(g_{n+1}f_{n-1}-g_{n-1}f_{n+1})-(f_{n+1}f_{n-1}-f_n^2)=0,
\end{equation}
where $D_t^na\cdot b=\partial_t-\partial_{t'})^n a(t)b(t')|_{t=t'}$ is the Hirota bilinear operator. The integrability is guaranteed if the bilinear equations admit $N$-soliton solution for every $N$ and arbitrary soliton parameters. However, it was observed that it is sufficient to have at least 3-soliton solution \cite{hir} to have integrability.

Accordingly, the $3$-soliton solution  is given by:
$$g_n=\sum_{\mu_1,\mu_2,\mu_3\in\{0,1\}}(b_1\mu_1+b_2\mu_2+b_3\mu_3)\exp\left(\sum_{i=1}^3\mu_{i}\eta_i+\sum_{i<j}^3\mu_i\mu_ja_{ij}\right),$$
$$f_n=\sum_{\mu_1,\mu_2,\mu_3\in\{0,1\}}\exp\left(\sum_{i=1}^3\mu_{i}\eta_i+\sum_{i<j}^3\mu_i\mu_ja_{ij}\right).$$
where $\eta_i=k_i n+\omega_i t, b_i=(e^{k_i}-1)/(e^{k_i}+1), \omega_{i}=2\sinh k_i, \exp{a_{ij}}=(e^{k_i}-e^{k_j})^2/(1-e^{k_i+k_j})^2.$

The discrete version of potential KdV can be obtained directly by discretizing the bilinear forms imposing gauge invariance \cite{alfred} (i.e.invariance with respect to a multiplication with an exponential $\exp{(an+bt)}$ for any $a$ and $b$) \cite{hir1}, \cite{hir2}, \cite{hir3}, \cite{cory1}, \cite{cory2}. When we discretise time $t$ we put $t\to m$ and for the time-derivative we consider $\partial_t f(n,t)\to \frac{1}{h}(f(n,m+h)-f(n,m))$ where $h$ is the discretisation step (we can put $t=mh$ to have step 1 as well). Accordingly the bilinear time-operator will turn into
$$D_t g_n\cdot f_n\to \frac{1}{h}(\wt{g}f-f\wt{g})$$ where, for simplicity, we use the notation, $\wt{f}=f(n,m+h),\wt{\ov{f}}=f(n+1,m+h), \un{f}=f(n-1,m) $ etc.
So, replacing the time derivative in (\ref{bil0}) and keeping gauge invariance we get:
\begin{equation}\label{dddbil0}
\wt{g}f-g\wt{f}=h(\wt{\ov{g}}\un{f}-\wt{\ov{f}}\un{g}).
\end{equation}
We modify also the second bilinear equation (\ref{bil1}) (otherwise the resulting systems would be of higher order) such that to keep gauge invariance:
\begin{equation}\label{dddbil1}
(\wt{\ov{g}}\un{f}-\wt{\ov{f}}\un{g})-(\wt{\ov{f}}\un{f}-f\wt{f})=0.
\end{equation}
Now if $u=g/f$, dividing both (\ref{dddbil0}) and (\ref{dddbil1}) by $f\wt{f}$ and calling $B_0=\wt{\ov{f}}\un{f}/f\wt{f}$ we obtain:
$$\wt{u}-u=h(\wt{\ov{u}}-\un{u})B_0,$$
$$(\wt{\ov{u}}-\un{u})B_0-(B_0-1)=0.$$
Solving for $B_0$ in the second equation and plug into the first one we obtain:
$$\wt{u}-u=h\frac{\wt{\ov{u}}-\un{u}}{1-\wt{\ov{u}}+\un{u}},$$
which is exactly the potential discrete KdV of H1-type in the ABS list \cite{adler} and the bilinear equations (\ref{dddbil0}) and (\ref{dddbil1}) are precisely the bilinear equations for H1 equation found by Hietarinta and Zhang \cite{jarmo}. 
The potential discrete KdV used in \cite{liu} can be easily obtained. Indeed, if we change the independent variables $m\to m-n, n\to m$ and scalings $u\to \frac{u}{2(p_1-p_2)}, h\to-\frac{p_1+p_2}{p_1-p_2}$ we obtain:
$$u_{n+1,m+1}-u_{n,m}=\frac{2(p_1+p_2)(u_{n+1,m}-u_{n,m+1})}{2(p_2-p_1)+u_{n+1,m}-u_{n,m+1}}$$
Multisoliton solution has been computed elegantly in \cite{jarmo} using casoratians. Howeverm we shall use the old formulation with exponentials since is much more amenable to fermionic extension. The solution has the same form as in the semidiscrete case:
$$g_n=\sum_{\mu_1,...,\mu_N\in\{0,1\}}(b_1\mu_1+...+b_N\mu_N)\exp\left(\sum_{i=1}^N\mu_{i}\eta_i+\sum_{i<j}^3\mu_i\mu_ja_{ij}\right),$$
$$f_n=\sum_{\mu_1,...,\mu_N\in\{0,1\}}\exp\left(\sum_{i=1}^N\mu_{i}\eta_i+\sum_{i<j}^3\mu_i\mu_ja_{ij}\right),$$
where $\eta_i=k_i n+\omega_i hm, b_i=(1+h)(e^{k_i}-1)/(e^{k_i}+1), e^{h\omega_i}=(h-e^{k_i})/e^{k_i}(he^{k_i}-1), \exp{a_{ij}}=(e^{k_i}-e^{k_j})^2/(1-e^{k_i+k_j})^2.$

\subsection{Continuous super-KdV}
In order to understand soliton dynamics let us remind seversal aspects of super-KdV \cite{mathieu}. If $u_n(t)$ and $\psi_n(t)$
are even and odd grassmann functions then the super-KdV equation in the ${\cal N}=1$ superspace $(x,\theta)$, is written in terms of the odd superfield $\Phi=\psi+\theta u$ and super-covariant derivative ${\cal D}=\partial_{\theta}+\theta\partial_x$, as:
$$\Phi_t+\Phi_{xxx}+3(\Phi {\cal D}\Phi)_x=0.$$
If $\Phi=2{\cal D}^3\log \tau$ (where $\tau=F+\theta G$ is a bosonic tau function, with $F$-bosonic and $G$-fermionic) then the Hirota bilinear form is:
\begin{equation}\label{sb1}
(D_t+D_{x}^3)G\cdot F=0,
\end{equation}
\begin{equation}\label{sb2}
(D_tD_x+D_x^4)F\cdot F-2(D_t+D_x^3)G\cdot G=0.
\end{equation}
This form is equivalent with the well known super-bilinear form:
$$(S_xD_t+S_x^7)\tau\cdot \tau=0,$$
where $S^N a\cdot b$ is the super-Hirota operator defined in \cite{carstea0}. However, we shall rather be interested in the bilinear system (\ref{sb1}) and (\ref{sb2}) since in the discrete setting supersymmetry is not manifest.
The 2 super-soliton solution is given by \cite{carstea0, carstea1}:
$$G=\zeta_1 e^{\eta_1}+\zeta_2 e^{\eta_2}+a_{12}e^{\eta_1+\eta_2}\left(\alpha_{12}\zeta_1+\alpha_{21}\zeta_2\right)$$
$$F=1+e^{\eta_1}+e^{\eta_2}+a_{12}e^{\eta_1+\eta_2}\left(1+\frac{2}{k_2-k_1}\zeta_1\zeta_2\right)$$
where $\zeta_i$ are odd grassmann parameters, $\eta_i=k_i x-k_i^3 t$, $a_{ij}=(k_i-k_j)^2/(k_i+k_j)^2$ and $\alpha_{ij}=(k_i+k_j)/(k_i-k_j)$. The main difference between the interaction of ordinary solitons and of super-solitons is the appearance of the dressing factor $\alpha_{ij}$ and the correction ${2}/{(k_2-k_1)}\zeta_1\zeta_2$. So, at every interaction the soliton parameter is dressed with the factor $\alpha_{ij}$
\section{Bilinear discrete super-KdV equation}
We propose the following bilinear form for (\ref{liu1}) and (\ref{liu2}):

\begin{equation}\label{bil01}
D_t\gamma_n\cdot f_n=(\gamma_{n+1}f_{n-1}-\gamma_{n-1} f_{n+1}),
\end{equation}
\begin{equation}\label{bil02}
D_t g_n\cdot f_n-(g_{n+1}f_{n-1}-g_{n-1} f_{n+1})+D_t\gamma_n\cdot \gamma_n-2\gamma_{n+1}\gamma_{n-1}=0,
\end{equation}
\begin{equation}\label{bil03}
(g_{n+1}f_{n-1}-g_{n-1} f_{n+1})-(f_{n+1}f_{n-1}-f_n^2)-\frac{1}{2}D_t\gamma_n\cdot \gamma_n+ \gamma_{n+1}\gamma_{n-1}=0,
\end{equation}
where $\gamma_{n}(t)$ is a grassmann fermionic (odd) tau function and $g_n(t), f_n(t)$ are grassmann bosonic (even) tau functions.
The action of Hirota bilinear operator on purely fermionic functions is the same as in the ordinary case but one has to take into account the anticommutativity, $D_t\gamma_n\cdot \gamma_n={\dot{\gamma_n}}\gamma_n-\gamma_n{\dot{\gamma_n}}=2{\dot{\gamma_n}}\gamma_n$.
Let $\psi_n(t)=\gamma_n(t)/f_{n}(t)$, $u_n(t)=g_n(t)/f_n(t)$, $f_{n+1}f_{n-1}/f_n^2=B_n(t)$ 
Dividing (\ref{bil01}),(\ref{bil02}), (\ref{bil03}) by $f^2$ we get
\begin{equation}\label{bbil1}
\psi_{n,t}=(\psi_{n+1}-\psi_{n-1})B_n,
\end{equation}
\begin{equation}\label{bbil2}
u_{n,t}-(u_{n+1}-u_{n-1})B_n+\psi_{n,t}\psi-\psi_{n+1}\psi_{n-1}B_n=0,
\end{equation}
\begin{equation}\label{bbil3}
(u_{n+1}-u_{n-1})B_n-(B_n-1)+\psi_{n+1}\psi_{n-1}B_n=0.
\end{equation}
Solving for $B_n$ in (\ref{bbil3}) and then introducing in (\ref{bbil1}) and (\ref{bbil2}) we will get precisely
(\ref{liu1}) and (\ref{liu2}) (with $t\to -t$).

The integrability of the bilinear system (\ref{bil01}), (\ref{bil02}) and (\ref{bil03}) is proved by computing the multisoliton solution. We define the following quantities:
$$a_{ij}=\left(\frac{e^{k_i}-e^{k_j}}{e^{k_i+k_j}-1}\right)^2,\quad \omega_i=2\sinh k_i,$$ $$\alpha_{ij}=\left(\frac{e^{k_i+k_j}-1}{e^{k_i}-e^{k_j}}\right), b_i=\left(\frac{e^{k_i}-1}{e^{k_i}+1}\right), \delta_{ij}=\frac{\alpha_{ij}}{b_i+b_j}, \eta_i=k_i n+\omega_i t.$$
The 2-supersoliton solution is ($\zeta_1$ and $\zeta_2$ are odd grassmannian parameters):
$$\gamma_n=\zeta_1 e^{\eta_1}+\zeta_2 e^{\eta_2}+a_{12}e^{\eta_1+\eta_2}(\alpha_{12}\zeta_1+\alpha_{21}\zeta_2)$$
$$g_n=b_1e^{\eta_1}+b_2e^{\eta_2}+(b_1+b_2)a_{12}e^{\eta_1+\eta_2}(1+\delta_{12}\zeta_1\zeta_2)$$
$$f_n=1+e^{\eta_1}+e^{\eta_2}+2a_{12}e^{\eta_1+\eta_2}(1+\delta_{12}\zeta_1\zeta_2)$$
So, the structure of the solution is similar to the continuous case displaying the dressing factor in the fermionic component. The dressing mechanism allows extension to the 3-supersoliton solution which has the following form:
$$
\gamma_n=\sum_{i=1}^3\zeta_i e^{\eta_i}+\sum_{i<j}a_{ij}e^{\eta_i+\eta_j}(\alpha_{ij}\zeta_i+\alpha_{ji}\zeta_j)+a_{12}a_{13}a_{23}e^{\eta_1+\eta_2+\eta_3}((\alpha_{12}\alpha_{13}\zeta_1+
$$
\begin{equation}\label{3ssa}
+\alpha_{21}\alpha_{23}\zeta_2+\alpha_{31}\alpha_{32}\zeta_3)+\zeta_1\zeta_2\zeta_3(\delta_{12}\alpha_{13}\alpha_{23}\alpha_{31}\alpha_{32}-\delta_{13}\alpha_{12}\alpha_{32}\alpha_{21}\alpha_{23}+\delta_{23}\alpha_{21}\alpha_{31}\alpha_{12}\alpha_{13})),
\end{equation}
$$g_n=\sum_{i=1}^3b_ie^{\eta_i}+\sum_{i<j}(b_i+b_j)a_{ij}e^{\eta_i+\eta_j}(1+\delta_{ij}\zeta_i\zeta_j)+a_{12}a_{13}a_{23}(b_1+b_2+b_3)e^{\eta_1+\eta_2+\eta_3}+$$
\begin{equation}\label{3ssb}
+(b_1+b_2+b_3)a_{12}a_{13}a_{23}e^{\eta_1+\eta_2+\eta_3}(\delta_{12}\alpha_{13}\alpha_{23}\zeta_1\zeta_2+\delta_{23}\alpha_{21}\alpha_{31}\zeta_2\zeta_3+\delta_{13}\alpha_{12}\alpha_{32}\zeta_1\zeta_3),
\end{equation}
$$f_n=\sum_{i=1}^3b_ie^{\eta_i}+\sum_{i<j}a_{ij}e^{\eta_i+\eta_j}(1+2\delta_{ij}\zeta_i\zeta_j)+a_{12}a_{13}a_{23}e^{\eta_1+\eta_2+\eta_3}+$$
\begin{equation}\label{3ssc}
+a_{12}a_{13}a_{23}e^{\eta_1+\eta_2+\eta_3}(2\delta_{12}\alpha_{13}\alpha_{23}\zeta_1\zeta_2+2\delta_{23}\alpha_{21}\alpha_{31}\zeta_2\zeta_3+2\delta_{13}\alpha_{12}\alpha_{32}\zeta_1\zeta_3).
\end{equation}

\subsection{A new form of semidiscrete super-KdV equation}
Strating from the bilinear equations we can construct another integrable form which although simpler has a more complicated interaction terms between supersolitons. If we remove the term $D_t\gamma\cdot \gamma$ from the (\ref{bil03}) and put different coefficients we will obtain:
\begin{equation}\label{bil11}
D_t\gamma_n\cdot f_n=(\gamma_{n+1}f_{n-1}-\gamma_{n-1} f_{n+1}),
\end{equation}
\begin{equation}\label{bil12}
D_t g_n\cdot f_n-(g_{n+1}f_{n-1}-g_{n-1} f_{n+1})-\frac{1}{2}D_t\gamma_n\cdot \gamma_n+\gamma_{n+1}\gamma_{n-1}=0,
\end{equation}
\begin{equation}\label{bil13}
(g_{n+1}f_{n-1}-g_{n-1} f_{n+1})-(f_{n+1}f_{n-1}-f_n^2)+ \gamma_{n+1}\gamma_{n-1}=0.
\end{equation}

The 3-supersoliton solution will be almost the same but with a more complicated dressing in the three soliton interaction:
$$\gamma_n=\sum_{i=1}^3\zeta_i e^{\eta_i}+\sum_{i<j}a_{ij}e^{\eta_i+\eta_j}(\alpha_{ij}\zeta_i+\alpha_{ji}\zeta_j)+a_{12}a_{13}a_{23}e^{\eta_1+\eta_2+\eta_3}((\alpha_{12}\alpha_{13}\zeta_1+$$
$$+\alpha_{21}\alpha_{23}\zeta_2+\alpha_{31}\alpha_{32}\zeta_3)+\zeta_1\zeta_2\zeta_3(\delta_{12}\alpha_{13}\alpha_{23}\alpha_{31}\alpha_{32}-\delta_{13}\alpha_{12}\alpha_{32}\alpha_{21}\alpha_{23}+\delta_{23}\alpha_{21}\alpha_{31}\alpha_{12}\alpha_{13})),$$
$$g_n=\sum_{i=1}^3b_ie^{\eta_i}+\sum_{i<j}(b_i+b_j)a_{ij}e^{\eta_i+\eta_j}(1+\delta_{ij}\zeta_i\zeta_j)+a_{12}a_{13}a_{23}(b_1+b_2+b_3)e^{\eta_1+\eta_2+\eta_3}+$$
$$+(b_1+b_2+b_3)a_{12}a_{13}a_{23}e^{\eta_1+\eta_2+\eta_3}(\Delta_{12}\alpha_{13}\alpha_{23}\zeta_1\zeta_2+\Delta_{23}\alpha_{21}\alpha_{31}\zeta_2\zeta_3+\Delta_{13}\alpha_{12}\alpha_{32}\zeta_1\zeta_3),$$
$$f_n=\sum_{i=1}^3b_ie^{\eta_i}+\sum_{i<j}a_{ij}e^{\eta_i+\eta_j}(1+2\delta_{ij}\zeta_i\zeta_j)+a_{12}a_{13}a_{23}e^{\eta_1+\eta_2+\eta_3}+$$
$$+a_{12}a_{13}a_{23}e^{\eta_1+\eta_2+\eta_3}(2\Delta_{12}\alpha_{13}\alpha_{23}\zeta_1\zeta_2+2\Delta_{23}\alpha_{21}\alpha_{31}\zeta_2\zeta_3+2\Delta_{13}\alpha_{12}\alpha_{32}\zeta_1\zeta_3),$$
where 
$$\Delta_{ij}=\delta_{ij}\left(\frac{(b_i+b_k)+(b_j+b_k)}{b_1+b_2+b_3}\right)|_{k\neq i,j}.$$

The nonlinear form of this new bilinear system is:
\begin{equation}\label{liun1}
\partial_t \psi_n=\frac{\psi_{n+1}-\psi_{n-1}}{1+u_{n-1}-u_{n+1}},
\end{equation}
\begin{equation}\label{liun2}
\partial_t u_n=\frac{u_{n+1}-u_{n-1}}{1+u_{n-1}-u_{n+1}}+\frac{(1+u_{n-1}-u_{n+1})(\psi_n\psi_{n+1}-\psi_n\psi_{n-1})+\psi_{n+1}\psi_{n-1}}{(1+u_{n-1}-u_{n+1})^2}.
\end{equation}
\subsection{Continuum limit}
It is interesting to see what is the continuum limit of this new semidiscrete super-system (\ref{liun1}), (\ref{liun2}). We consider the following continuous variables:
$$x=\epsilon(n+2t), \tau=\frac{1}{3}\epsilon^3t, u_n(t)=\frac{1}{4}\epsilon W(x,\tau), \psi_n(t)=\frac{1}{2}\epsilon^{1/2}\Phi(x,\tau).$$
Everything will cancel up to the order $\epsilon^4$ where we get the following system:
$$W_{\tau}-3W_x^2-W_{xxx}-3\Phi_x\Phi_{xx}-3\Phi_x\Phi W_x=0,$$
$$\Phi_{\tau}-3\Phi_xW_x-\Phi_{xxx}=0,$$
which looks like a new fermionic extension of KdV equation. However this is not so and in fact it is equivalent to the well known supersymmetric KdV system up to a transformation $W_x\rightarrow W_x+\Phi_x\Phi$. So practically we do have {\it two integrable discretisations} of supersymmetric KdV.

\begin{remark} If $m_1=m_2, m_3=m_2+m_4$ then the following general bilinear system:
\begin{equation}\label{gbil1}
D_t\gamma_n\cdot f_n=(\gamma_{n+1}f_{n-1}-\gamma_{n-1} f_{n+1}),
\end{equation}
\begin{equation}\label{gbil2}
D_t g_n\cdot f_n-(g_{n+1}f_{n-1}-g_{n-1} f_{n+1})-\frac{m_1}{2}D_t\gamma_n\cdot \gamma_n+m_2\gamma_{n+1}\gamma_{n-1}=0,
\end{equation}
\begin{equation}\label{gbil3}
(g_{n+1}f_{n-1}-g_{n-1} f_{n+1})-(f_{n+1}f_{n-1}-f_n^2)+\frac{m_3}{2}D_t\gamma_n\cdot \gamma_n+ m_4\gamma_{n+1}\gamma_{n-1}=0,
\end{equation}
has {\it always} two-supersoliton solution in the form:
$$\gamma_n=\zeta_1\gamma_n^1+\zeta_2\gamma_n^2,\quad g_n=g_n^0+\zeta_1\zeta_2 g_n^1,\quad f_n=f_n^0+\zeta_1\zeta_2 f_n^1$$
where:
$$\gamma_n^1=e^{\eta_1}+\alpha_{12}a_{12}e^{\eta_1+\eta_2},\quad \gamma_n^2=e^{\eta_2}+\alpha_{21}a_{12}e^{\eta_1+\eta_2}$$
$$g_n^0=b_1 e^{\eta_1}+b_2 e^{\eta_2}+(b_1+b_2)a_{12}e^{\eta_1+\eta_2}, \quad g_n^1=-m_2 a_{12}\delta_{12}(b_1+b_2)e^{\eta_1+\eta_2}$$
$$f_n^0=1+e^{\eta_1}+e^{\eta_2}+a_{12}e^{\eta_1+\eta_2}, \quad f_n^1=2m_4a_{12}\delta_{12}e^{\eta_1+\eta_2}.$$
\end{remark}
\section{Time discretization}

In this section we use the semi-discrete superbilinear forms to construct integrable time discretization of the above equations. The main advantage is that bilinear equations can be discretized straightforwardly. Some difficulties appear at recovering nonlinear forms. As we said in the section 2 the guiding principle in discretizing the bilinear equations is the so called gauge invariance \cite{alfred} which means that any bilinear equation to be invariant with respect to a multiplication with an exponential of any linear combination of $x$ and $t$. 

Using this procedure we discretise directly the bilinear system (\ref{gbil1}), (\ref{gbil2}), (\ref{gbil3}). We find (keeping the same notations as in the section 2 and imposing gauge-invariance i.e the number of tildes on any term to be the same):
\begin{equation}\label{gdbil1}
\wt{\gamma}f-\gamma\wt{f}=h(\wt{\ov{\gamma}}\un{f}-\un{\gamma}\wt{\ov{f}}),
\end{equation}
\begin{equation}\label{gdbil2}
\wt{g}f-g\wt{f}-h(\wt{\ov{g}}\un{f}-\un{g}\wt{\ov{f}})-m_1\wt{\gamma} \gamma+h m_2\wt{\ov{\gamma}}\un{\gamma}=0,
\end{equation}
\begin{equation}\label{gdbil3}
h(\wt{\ov{g}}\un{f}-\un{g}\wt{\ov{f}})-h(\wt{\ov{f}}\un{f}-f\wt{f})+m_3\wt{\gamma}\gamma+ h m_4\wt{\ov{\gamma}}\un{\gamma}=0.
\end{equation}

For $m_1=m_2, m_3=m_2+m_4$ the above system admits 2-supersoliton solution. However for $m_2=-2, m_4=1$ the above system admits 3-supersoliton solution. It has the same form as (\ref{3ssa}),(\ref{3ssb}), (\ref{3ssc}) provided we redefine the quantities as follows:
$$\eta_i=k_i n+\omega_i h t, b_i=(1+h)(e^{k_i}-1)/(e^{k_i}+1), e^{h\omega_i}=(h-e^{k_1})/e^{k_1}(he^{k_1}-1)$$
Also the case of $m_1=-1, m_3=0$ is integrable and have three soliton solution with the same form as above except $\delta_{ij}\to \Delta_{ij}$

The nonlinear form of the bilinear system with $m_4=1, m_2=-2$ can be constructed quite easily in the same way as in the case of differential-difference version namely:

\begin{equation}\label{dliu1}
\wt{\psi}-\psi=h\frac{\wt{\ov{\psi}}-\un{\psi}}{1-\wt{\ov{u}}+\un{u}},
\end{equation}
\begin{equation}\label{dliu2}
\wt{u}-u=h\frac{\wt{\ov{u}}-\un{u}}{1-\wt{\ov{u}}+\un{u}}+h\frac{2-\wt{\ov{u}}+\un{u}}{(1-\wt{\ov{u}}+\un{u})^2}(\wt{\ov{\psi}}-\un{\psi})(\psi-\un{\psi}),
\end{equation}
which is precisely the form (\ref{diliu1}) and (\ref{diliu2})(up to the transformation of the independent variables shown in section two)
Also in the case of $m_2=-1$ we get the following nonlinear form:
\begin{equation}\label{ddliu1}
\wt{\psi}-\psi=h\frac{\wt{\ov{\psi}}-\un{\psi}}{1-\wt{\ov{u}}+\un{u}},
\end{equation}
\begin{equation}\label{ddliu2}
\wt{u}-u=h\frac{\wt{\ov{u}}-\un{u}}{1-\wt{\ov{u}}+\un{u}}+h\frac{(1-\wt{\ov{u}}+\un{u})\psi(\wt{\ov{\psi}}-\un{\psi})+\wt{\ov{\psi}}\un{\psi}}{(1-\wt{\ov{u}}+\un{u})^2}.
\end{equation}

\section{Conclusions}

What is really new in this paper is that we were able to compute soliton solution for a discrete integrable super-equation which is not supersymmetric. Usually supersymmetry has a huge role in simplifying computations. Non-supersymmetric extensions of integrable equations are much more complex and we expected that the solutions and bilinear forms to be as well. However, due to the particular form of the semidiscrete super-system (\ref{liu1}) and (\ref{liu2})which comes from a B\"acklund transformation of a supersymmetric continuous one, the solutions were rather easily found. The solitons look practically the same as in the continuous case except the second case where the interaction is more complicated. In addition, the fact that we found two different discretisations shows that nonlinear integrable interaction of bosonic and fermionic fields in a discrete world can be extremely intricate.


\begin{thebibliography}{100}
\bibitem{liu} L. L. Xue, D. Levi, Q. P. Liu, J. Phys. A: Math. Theor. 46, 502001, (2013)
\bibitem{aratyn}{\it Supersymmetric Integrable Models} in Lect. Not. in Physics, 502, (1998), Springer.
\bibitem{manin}Y. Manin, A. Radul, Comm. Math. Phys. 98, 65, (1985)
\bibitem{kac1}V. Kac, J. van de Leur, Ann. Inst. Fourier, 37, 99, (1987)
\bibitem{kac2}V. Kac, E. Medina, Lett. Math. Phys. 37, 435, (1996)
\bibitem{leclair}A. LeClair, Nucl. Phys. B, 314, 435, (1989)
\bibitem{ueno}K. Ueno, H. Yamada, K. Ikeda, Comm. Math. Phys. 124, 57, (1989)
\bibitem{carstea0}A. S. Carstea, Nonlinearity 13, 1645, (2000)
\bibitem{carstea1}A. S. Carstea, A. Ramani, B. Grammaticos, Nonlinearity 14, 1419, (2001)
\bibitem{carstea2}B. Grammaticos, A. Ramani, A. S. Carstea, J. Phys. A: Math. Gen. 34, 4881, (2001) 
\bibitem{carstea3}A. S. Carstea, A. Ramani, B. Grammaticos, Chaos, Solitons and Fractals, 14, 155, (2002)
\bibitem{carstea4}A. Ramani, A. S. Carstea, B. Grammaticos, Phys. Lett. A292, 115, (2001)
\bibitem{ssh1}A. S. Carstea, D. Grecu, A. Visinescu, Europhysics Letters, 67, 531, (2004)
\bibitem{ssh2}A. S. Carstea, Chaos, Solitons and Fractals, 42, 923, (2009)
\bibitem{hu1}Q. P. Liu, X. B. Hu, M. X. Zhang, Nonlinearity 18, 1597, (2005); 
\bibitem{hu2}Q. P. Liu, X. B. Hu, J. Phys. A: Math. Gen. 38, 6371, (2005)
\bibitem{c1}Q. P. Liu, X. X. Yang, Physics Lett. A351, 131, (2006)
\bibitem{c2}E. Fan, Stud. Appl. Math. 127, 284, (2011)
\bibitem{mathieu}P. Mathieu, J. Math. Phys. 2499, (1988)
\bibitem{grakhovski}G. G. Grahovski, A. V. Mikhailov, arXiv:1303.1853, Phys. Lett. A to appear. 
\bibitem{hirota}R. Hirota, J. Satsuma, Progr. Theor. Physics, 59, 64, (1976)
\bibitem{alfred}B. Grammaticos, A. Ramani, J. Hietarinta, Phys. Lett. A190, 65, (1994)
\bibitem{jarmo}J. Hietarinta, Da-jun Zhang, J. Phys. A: Math. Theor. 42, 404006, (2009)
\bibitem{hir}J. Hietarinta, J. Math. Phys. 28, 2094, (1987); 28, 2586, (1987)
\bibitem{adler}V. Adler, A. Bobenko, Yu. Suris, Comm. Math. Phys. 233, 513, (2003)
\bibitem{hir1}R. Hirota, {\it Discretization of soliton equations}, lecture presented at SIDE III: Symmetries and Integrability of Difference Equations, Sabaudia, Italy, 1998
\bibitem{hir2}R. Hirota, Chaos, Solitons and Fractals, 11, 77, (2000)
\bibitem{hir3}R. Hirota in {\it Bilinear Integrable Systems: From classical to quantum, continuous to discrete}, L. Fadeev, P. van Moerbecke, F. Lambert Eds. 113-122,(2006) 
\bibitem{cory1}C. Babalic, A. S. Carstea, J. Phys. A: Math. Theor. 46, 145205, (2013)
\bibitem{cory2}C. Babalic, A. S. Carstea, Cent. Eur. J. Physics, 12, 341, (2014)
\end{thebibliography}
\end{document}